# Triggering the Continuous Growth of Graphene toward Millimeter Size Grain

Tianru Wu[a,b], Guqiao Ding[b], Honglie Shen[a*], Haomin Wang[b], Lei Sun[a], Da Jiang[b], Xiaoming Xie[b*], Mianheng Jiang[b]

[a]*College of Materials Science & Technology, Nanjing University of Aeronautics & Astronautics, Nanjing, 211100, China*

[b]*State Key Laboratory of Functional Materials for Informatics, Shanghai Institute of Microsystem and Information Technology, Chinese Academy of Sciences, 865 Changning Road, Shanghai 200050, People's Republic of China*

*Correspondence to: Xiaoming Xie (xmxie@mail.sim.ac.cn), Honglie Shen (hlshen@nuaa.edu.cn)*

**Abstract**

In this report, we demonstrated a simple but efficient strategy to synthesize millimeter-sized graphene single crystal grains by regulating the supply of reactants in chemical vapor deposition process. Polystyrene was used as a carbon source. Pulse heating on the carbon source was utilized to minimize the nucleation density of graphene on copper foil, while the gradual increase in the temperature of carbon source and the flow rate of hydrogen is adapted to drive the continuous growth of graphene grain. As a result, the nucleation density of graphene grain can be controlled as lower as ~100 nuclei/cm$^2$, and the dimension of single crystal grain could grow up to ~1.2 mm. Raman spectroscopy, transmission electron microscopy and electrical transport measurement show that the graphene grains obtained are in high quality. The strategy presented here provides very good controllability and enables the possibility for large graphene single crystals, which is of vital importance for practical applications.

## 1. Introduction

Graphene, a monolayer of sp$^2$ carbon atoms, has been attracting great interests as an ideal two dimensional (2D) crystalline material.[1] Its 2D nature and high carrier mobility provide a great platform for exploration of high performance electronics. Fabrication techniques for wafer scale graphene are already available. However, graphene in wafer scale differs from normal silicon as it is mainly polycrystalline with typical grain size in the range of microns or tens of microns.[2-4] To synthesis continuous graphene film with large grain size is still a great challenge.[5-6]

Synthesis of graphene with millimeter-sized grains was demonstrated by several groups. Wang *et al.*[7] realized the growth of rectangular-shaped graphene domain on copper foil with dimension of about 0.4mm based on ambient pressure CVD (APCVD). They found that pre-annealing of Cu foil could remarkably reduce the graphene nucleation density. Li *et al.*[8] succeeded in making graphene domain with lateral size of 0.5 mm, via low pressure chemical vapor deposition (LPCVD) process by using methane as a precursor at 1035 $^o$C. The graphene grains obtained are dendrite-shaped, which may contain single-crystal-subgrains. They found that graphene growth rate gradually slows down with the increase of growth time, due to reduced copper surface exposure. Later, Gao *et al.*[9] achieved millimeter-sized grain in hexagonal shape on Pt foil. It took more than 3 hours to grow ~1 mm single-crystal grains. Although the above experiments demonstrated the growth of large grains, none of them gave a clear picture about the growth dynamics.

In this manuscript, we demonstrated a simple but efficient strategy to synthesis millimeter-sized graphene grains by APCVD. The key is to maintain low nucleation rate and to provide continuous drive for graphene growth. Nucleation density as low as ~100 nuclei/cm$^2$ was achieved by optimizing process parameters on copper substrate subjected to fine polishing and pre-annealing. A continuous increase of carbon supply and the total gas flow rate drive the continuous graphene growth. Single crystal graphene domain of 1.2 mm in size was successfully synthesized, which is the best result synthesized on Cu substrate published so far via the strategy.

## 2. Results and Discussion:

Polystyrene was chosen as the carbon source because of its relative weak C-H bonds and low decomposition temperature compared to the widely used gaseous sources. We heated the polystyrene to a temperature lower than 280 $^o$C (defined as T$_p$ hereafter). The products decomposed from polystyrene were analyzed by means of gas chromatography and mass spectroscopy. The results are shown in Figure 1a and 1b, respectively. From Figure 1b, it is found that polystyrene starts to decompose at about 80 $^{o}$C. The product is composed of various benzene-ring containing radicals. The data are consistent to that of the earlier literature.[10] The radicals are guided to the growth zone in the furnace by the carrier gases, which is a mixture of Ar/H$_2$. Under high temperature, the radicals are dehydrogenated, and then connect with each other to form graphene. The schematic of CVD growth setup is depicted in Figure S1 in the supporting information.

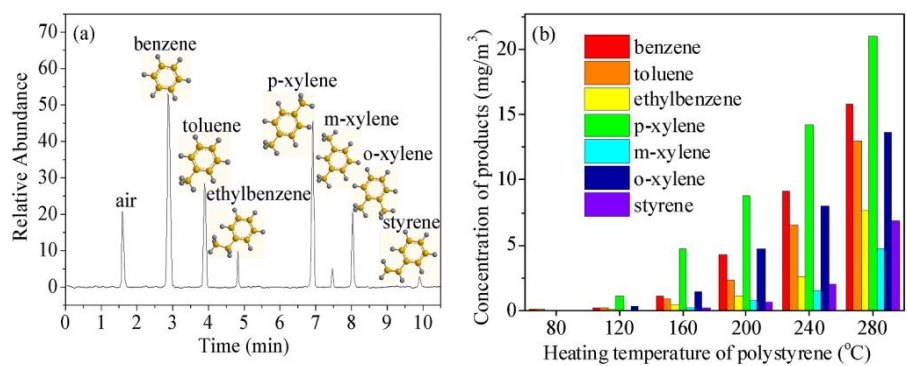

**Figure 1.** Investigation of polystyrene decomposition at different temperatures. (a) Gas chromatography of polystyrene decomposed at 280 $^{o}$C; (b) Concentration of products decomposed from polystyrene at different temperatures.

To realize low nucleation density is the prerequisite for growing large single crystal graphene grains. Defects on the substrate, such as impurities, dislocation, surface irregularities normally serve as nucleation centers (see Figure S2). To reduce its contribution, the copper foil was fine polished and annealed at 1050 $^{o}$C in Ar/H$_2$ mixture for 60 minutes prior to graphene growth. The surface treatment was proven to be an important step for the controllable graphene growth. Beside substrate defects, we found that H$_2$ flow rate also plays an important role on graphene nucleation.[11] Systematic investigations on the influence of H$_2$ flow rate was carried out. The growth temperature was kept at 1050 $^{o}$C, and T$_p$ maintained at 230 $^{o}$C. The dependence of nucleation density on H$_2$ flow rate was plotted in Figure 2a. The nucleation density stays at a low value (around several hundred per centimeter square) when H$_2$ flow rate was lower than 2 sccm, it increases dramatically at higher flow rates by more than three orders of magnitude. It is necessary to mention that H$_2$ flow rate lower than ~2 sccm yields dendrite-shaped nucleus (as shown in the upper insert of Figure 2a) due to high anisotropic growth, while high flow rate results in hexagonal shaped nucleus (as shown in the lower insert of Figure 2a).The effects of hydrogen on nucleation density and grain shape are consistent with early observations which revealed hydrogen's double roles[12] as an activator of carbon related radicals that promotes to the graphene nucleation, and as an etching reagent that controls the shape of the resulting graphene grains. Proper active hydrogen radicals catalyzed on Cu foil could promote the activation of cyclobenzene related molecules by dehydrogenation effect, and the threshold of nucleation was promoted due to the supersaturation of more activated carbon related radicals on Cu foil.

In order to further optimize the behavior of graphene nucleation, we decided to examine the effect of $T_p$ on the formation of graphene nuclei at different $H_2$ flow rate. Carbon supersaturation determined by $T_p$ will certainly affects the nucleation. For nucleation of graphene, it has to overcome an additional energy barrier ($\Delta G^*$) to form crystal nucleus, companied by a decrease of free energy which occurs spontaneously.[13] The energy barrier does not exist for graphene grain growth. Nucleation threshold at different hydrogen flow rates were studied and the critical $T_p$ at each hydrogen flow rate was obtained as shown in Figure 2b. The results suggest that a hydrogen flow rate ~ 2 sccm and $T_p$ ~205 °C would be a very good set of process parameters for low nucleation density and control the hexagonal shape of the graphene nucleus.

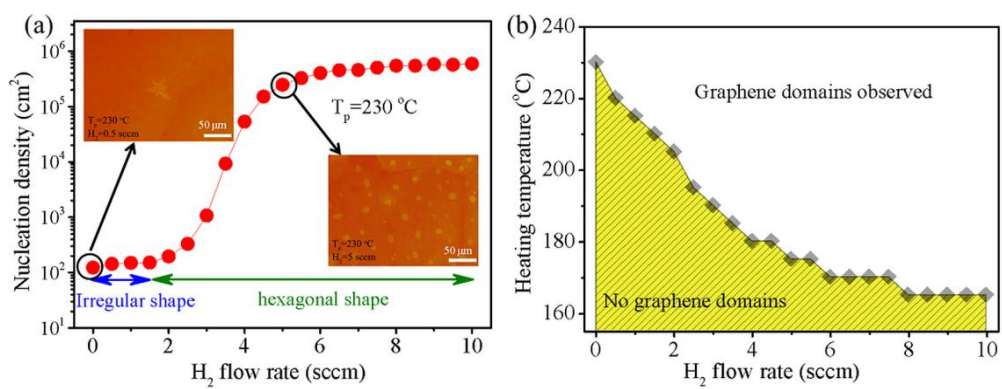

**Figure 2.** (a) The dependence of the nucleation density on the $H_2$ flow rate from 0 to 10 sccm. The inserts show optical images of graphene grains synthesized under different $H_2$ flow rates at 1050 °C for 2 min with $H_2$ flow at 0.5sccm, and 5 sccm, $T_p$=230°C; (b) Critical temperature of nucleation as a function of the $H_2$ flow rate.

From above experiments, we developed the technology to achieve a lower nucleation density. In order to achieve large graphene grain, it is important to drive the graphene growing at a comparatively high rate. Normally, higher growth temperature is evidently beneficial for graphene grain growth (discussed in the supporting information and Figure S3). Apparently, this is due to the enhancement of the catalysis of the substrate metal at higher temperatures, which enables the higher speed of carbon construction at the edge of graphene grain. Nevertheless, the present investigations also indicate the temperature shall not be too close to the melting point of the substrate, which will produce excessive defects in the graphene domains. According to above discussion, it should be necessary to choose a proper temperature for growth ( high but not too close to the melting point of the substrate) and maintain a low $H_2$ flow rate (yet high enough to have sufficient etching power to control the graphene shape) to make graphene grain grow as large as possible.

Our strategy to grow the graphene grain in millimeter is clear. If the nuclei density is low enough, gradual increase in the supply of polystyrene is adapted to drive the continuous growth of graphene toward millimeter. We set $T_p$ to the critical temperature transiently to yield a low density of graphene nuclei on copper foil, and then gradually increased $T_p$ and reactant gas flow rate to drive the continuous growth of graphene. The optimized process is shown in Figure 3. At the nucleation stage (as shown in the pink area in Figure 3a), the $H_2$ flow rate was set at 2 sccm while keeping the total flow rate at 300 sccm. $T_p$ has to increase to a relatively high value (205 $^o$C) to achieve supersaturation of activated carbon related radicals for initiating the nucleation (shown in Figure 3b). Once the graphene crystal nucleuses were formed, $T_p$ was decreased to below the nucleation threshold so that further nucleation was largely inhibited. During the growth stage, continuous ramping in $T_p$ and total gas flow rate was employed. $T_p$ increased from 195 $^o$C to 215 $^o$C and the total gas flow increased for from 300 sccm to 600 sccm for a total time of 80 min, with the partial pressure of $H_2$ kept at the same value. The grain growth at increasing supply of active carbon related radicals overcomes the etching effect by $H_2$, resulting in a continuous growth of graphene grains. Compared to normal process (Figure 3c), larger grain size of ~500 m was achieved after 40 min due to higher growth rate (Figure 3d). It is interesting to note that if we apply a stepwise increase of $T_p$, we could obtain large graphene domains with stripe-like edges (Figure S4).

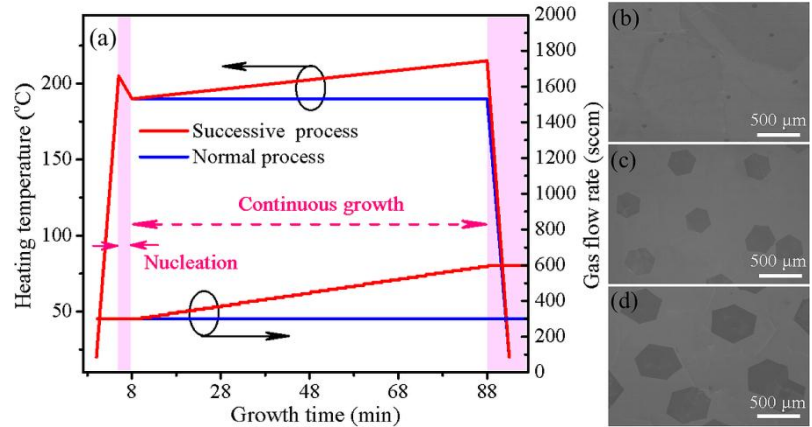

**Figure 3.** (a) APCVD growth of graphene at 1050 °C with linear increase of $T_p$ and total reactant gas flow rate; (b) SEM image of graphene nucleation site; (c) SEM image of hexagon-shaped graphene under normal growth process for 40 min; (d) SEM image of hexagon-shaped graphene under continuous growth process for 40 min.

It was found that the growth rate of graphene grain gradually decreases with the increase of the grain size after 40 min by normal process (Figure 4e). Sometime, the graphene grains do not grow any more when reaction equilibrium is gradually established at the edge of graphene grains. The concentration of active carbon related radicals is not high enough to maintain the growth rate because of the reduction of the copper catalytic effect.[14] The catalytic effect of degrades due to the reduction of exposed Cu surface area. To increase the grain size further, it is necessary to break the thermodynamic equilibrium of graphene growth. As shown in Figure 4e, It is clear that the process of increasing $T_p$ and total gas flow rate can yield continuous growth of graphene grains even at 80 min. In order to find the optimal condition and to separately investigate the effect of $T_p$ and the total gas flow rate, we conducted a series of experiments (as shown in Figure 4e). By normal process, the grain size gradually saturates at about ~420 μm at 80min. Increasing the gas flow rate alone from 300 sccm to 600 sccm obtained the grain size to ~630 μm, while increasing $T_p$ alone from 195 °C to 215 °C yields in a grain size of ~750 μm. If $T_p$ and the gas flow rates are both increased, a much higher growth rate was observed, and the maximum domain size reached 1.2 mm, the biggest one reported on copper substrate. Figure 4a-4d show optical images of hexagonal graphene grain synthesized in 20 min, 40 min, 60 min, 80 min, respectively. It was also proved that the grain boundaries of Cu foil do not restrain the continuous growth of graphene domains. Individual hexagonal graphene grains can grow continuously across Cu grain boundaries without any apparent distortion, as shown in Figure 4. This phenomenon reflects the weak influence of the Cu crystal lattice on graphene growth and demonstrates that single-crystalline graphene can grow on polycrystalline Cu foil. Compared to normal process, it is clear that to gradually increase $T_p$ and reactant gas flow rate could drive the grain grow continuously. The possible reason is that 1) the continuous ramping in $T_p$ increases the density of active carbon related radicals on Cu foil and breaks the dynamic equilibrium at graphene edges. The process results in the continuous growth of graphene grain; 2) the increasing gas flow rate also enhances the feed rate and movement of active carbon related radicals and the removal of unwanted by-product from Cu surface.[15] Both of the process parameters drive graphene grain toward larger size. Compared to the two-step process,[16] our strategy has obvious advantage in obtaining graphene in large size.

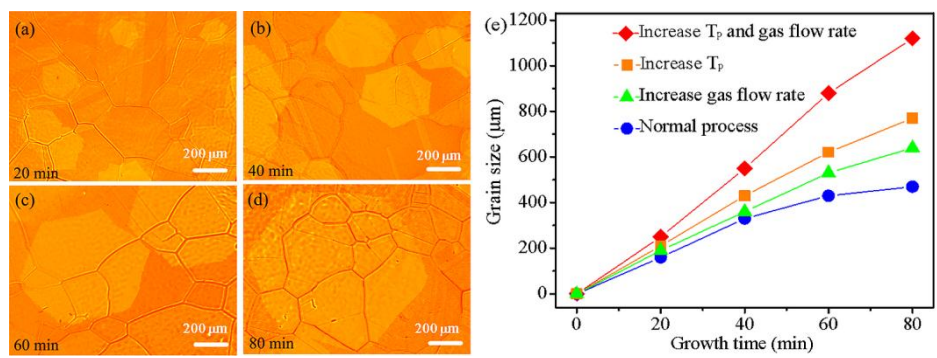

**Figure 4.** Optical images of hexagonal graphene grains synthesized for (a) 20 min, (b) 40 min, (c) 60 min, and (d) 80 min; (e) The grain size as the function of growth time under different process.

As Raman measurement could provide us the information about the number of layer and defect densities of individual hexagonal graphene domain, we conduct Raman measurement on a hexagonal graphene grain (Figure5a), which was transferred onto $SiO_2$/Si substrate. The result was shown in Figure 5b and 5c. It is found that the full width at half maximum of 2D peak is about 28-35 cm$^{-1}$, the intensity ratio between 2D and G peaks ($I_{2d}/I_G$) is 2-3. The D peak is not observed (Figure 5d). The results indicate that most of isolated hexagon-shaped graphene grains were single layer graphene. Meanwhile, transmission electron microscopy (TEM) of hexagonal graphene grains was introduced for crystalline structure characterizations of millimeter sized graphene domains (Figure 5e and 5f). The high-resolution TEM image (Figure 5g) indicates the graphene grain is monolayer. Selective area electron diffraction (SAED) data on different positions show the same set of hexagonal diffraction spots without rotation. It indicates that the detected area is single crystalline (Figure 5h).

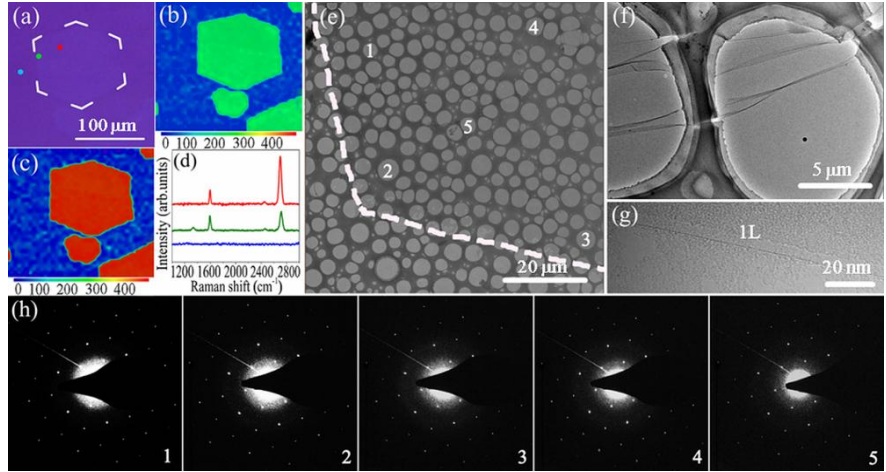

**Figure 5.** (a) Optical image of hexagonal graphene grains on SiO$_2$/Si substrate; (b) Intensity maps of the G bands; (C) Intensity maps of the 2D bands; (d) Raman spectra taken in the circled area shown in (a); (e) Low magnification TEM image of 120 °C corner graphene grain transferred to TEM grid; (f) TEM image of graphene wrinkle on the grid; (g) The high-resolution TEM image of monolayer gaphene; (h) SAED for five regions indicated 1 to 5 shown in (e).

To investigate the electronic properties of samples synthesized at 1050 °C, the graphene-based back-gate Łeld-effect transistors (FETs) were fabricated on 300 nm SiO$_2$/Si substrates (see Figure S5). A typical channel width and length were 5 m and 10 m. Figure 6a shows the typical source-drain current ($I_{ds}$) vs the source-drain voltage ($V_{sd}$) curves at different gate voltage ($V_g$). The linear $I_{ds}$-$V_{ds}$ dependence in both cases reveals good ohmic contact between Cr/Au pads and graphene layer. The field effect mobility (μ) of graphene sheets can be achieved by using the equation:

$$R_{total} = \frac{V_{sd}}{I_{sd}} = \frac{L}{W\mu\sqrt{(n_oe)^2 + (V_g - V_{diac})^2 C_g^2}} + R_c \quad (1)$$

Where $R_{total}$ is the device resistance including the channel resistance and contact resistance $R_c$,[17] $e$ is the electron charge, and $n_0$ is the carrier density due to residual impurities, respectively.[18] $C_g$ is the gate capacitance per unit area (11 nF·cm$^{-2}$), and $V_{sd}$ is source-drain voltage (0.1V). $L$ and $W$ are channel length and width, respectively. The samples were monitored by $I_{ds}$-$V_g$ curves in the measurement at room temperature. Histogram of the field effect mobility distribution for total 22 millimeter sized graphene domains is about 5000-8000 cm$^2$V$^{-1}$s$^{-1}$, suggesting that the samples synthesized by polystyrene are of high quality.

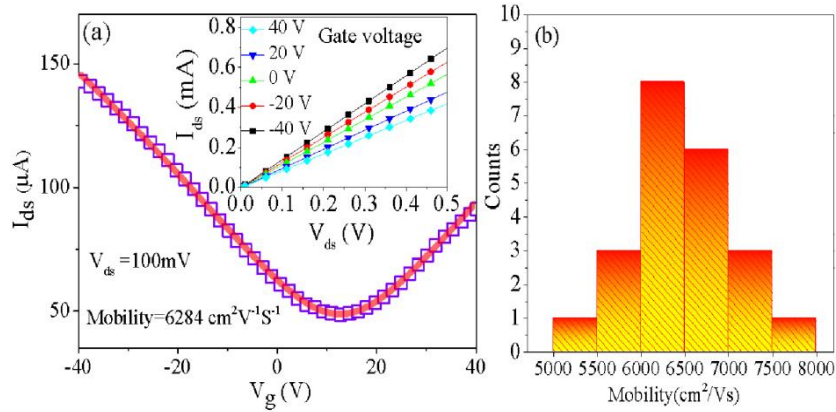

**Figure 6.** (a) Electrical characteristics ($I_{ds}$-$V_g$) of devices at $V_{ds}$=100 mV. The inset is $I_{ds}$-$V_{ds}$ characteristics at various $V_g$ for the graphene devices we fabricated; (b) Histogram of the field effect mobility distribution for total 22 devices.

## 3. Summary

In this study, we demonstrated a simple but efficient strategy to synthesis millimeter graphene grain by controlling the supply of carbon source in chemical vapor deposition process. Polystyrene was used as a carbon source, which could provide active benzene-ring radicals. The density of the graphene nucleation and the growth rate of graphene grain can be controlled by tuning the temperature of polystyrene and $H_2$ flow rates. By optimizing process parameters, nucleation density as lower as ~100 nucleus/cm$^2$ was achieved, and hexagonal shaped single crystal grain up to 1.2 mm was obtained, with field effect mobility ranging from 5000-8000 cm$^2$V$^{-1}$s$^{-1}$. The Raman measurement, TEM results and electrical transport data indicates that graphene are of high quality. With nucleation density of ~100 nuclei/cm$^2$, the average distance between nuclei is already in millimeter range, so in order to increase further the single crystal grain size, lower nucleation density is required which demands for further engineering efforts. Our strategy offers a high controllability of graphene growth with millimeter-size, and will definitely benefit for the practical application in graphene based electronics.

## 4. Experimental Section

Copper foil (Maikun Chemical 99.9 %) was used as catalytic substrate. The foil was first mechanically and electrochemically polished, and then subjected to annealing at 1050 °C for 60 min in a mixture of H$_2$ and Ar (1:10) with gas flow rate of 300 sccm. After the pretreatment, smooth Cu surface with 0.5-3 mm large grain size could be obtained. The polishing and annealing processes are very important in reducing the vacancies, dislocations, stacking faults, phase or grain boundaries, impurity atoms and surface irregularities.[19] These processes could effectively reduce the nucleation centers. Polystyrene (Aladdin MW 100,000) was used as solid carbon source. It was loaded in a small one-side-sealed container, placed at the gas influx side of the quartz tube in the growth chamber.

The graphene synthesis was done by APCVD at a typical temperature ranging 950-1050 $^o$C for 30-80 min. A mixture of Ar and H$_2$ are used as reactant gases. The H$_2$ flow rate was adjustable between 0-10 sccm while keeping the total flow rate at 300 sccm. Carbon supply was controlled by heating the precursor with a halogen lamp under typical polystyrene temperature (T$_p$) range of 80-280 $^o$C. After the growth process, the halogen tungsten lamp was turned off, and the furnace was cooled down to room temperature naturally.

Characterizations were done by optical microscopy (Leica material Microscope DM6000M), Raman microprobe spectroscopy (Thermo Fisher DXR) using an Ar$^+$ laser (wavelength 532 nm) with 1 m laser spot, equipped with an optical microscopy, scanning electron microscopy (SEM, FEI NOVA Nano SEM) with operating voltage of 5 kV were used for morphological investigation. Gas chromatography (Agilent 7890A-5975C) was used to investigate the products of polystyrene during thermal decomposition. The crystalline of the hexagonal grains was characterized by Transmission Electron Microscope (TEM, Philips CM-200 FEG, operated at 200 kV).

**Acknowledgements**

This work was supported by projects from the National Natural Science Foundation of China (Grant No. 11104303 and 61136005), the National Science and Technology Major Project (Grant No. 2011ZX02707), Shanghai Science and Technology Commission (Grant No. 10DJ1400600) and Priority Academic Program Development of Jiangsu Higher Education Institutions.

Supporting information

**To Trigger the Continuous Growth of Graphene toward Millimeter Size Grain**

Tianru Wu[a,b], Guqiao Ding[b], Honglie Shen[a*], Haomin Wang[b]，Lei Sun[a], Da Jiang[b], , Xiaoming Xie[b*], Mianheng Jiang[b]

[a]*College of Materials Science & Technology, Nanjing University of Aeronautics & Astronautics, Nanjing, 211100, China*

[b]*State Key Laboratory of Functional Materials for Informatics, Shanghai Institute of Microsystem and Information Technology, Chinese Academy of Sciences, 865 Changning Road, Shanghai 200050, People's Republic of China*

**1) Graphene growth by CVD using polystyrene as a solid carbon source**

Polystyrene was chosen as the solid carbon source, because it decomposes easily at temperatures in 100-300 $^oC$, allowing for easy controllability. We designed a setup to synthesis CVD graphene by using polystyrene as sold carbon source (Figure S1). Graphene can be synthesized at various temperatures (950 $^oC$-1050 $^oC$).

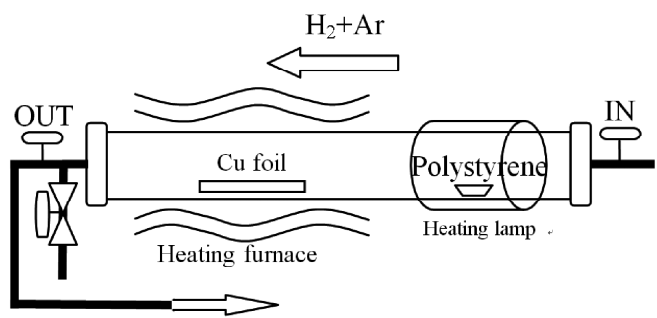

**Figure S1.** The setup for CVD growth of graphene.

**2) Cu substrate pretreatment**

Graphene prefers to nucleate at defect sites on the surface.[1] Figure S2a shows excessive graphene nucleation along stripes due to cold rolling process. The nucleation density can be greatly reduced by polishing and high temperature annealing, which removes the uncontrollable defects on the substrate surface.[2]

However, if polishing was not done very well, abrasive particles may be embedded in the copper matrix, and becoming preferred nucleation sites (see Figure S2b). In our work, mechanical polishing, electrochemical polishing and annealing process (1050°C) were used to obtain smooth Cu surface with roughness around 3 nm and large grain size in between 0.2-3 mm. With high temperature annealing of polished Cu foil, the vacancies, dislocations, stacking faults, and grain boundaries are reduced before the synthesis of graphene.[3] The flat Cu surface can also enlarge the diffusion length of the carbon atoms. Both these factors lead to a reduction of the graphene domain density at a high temperature.

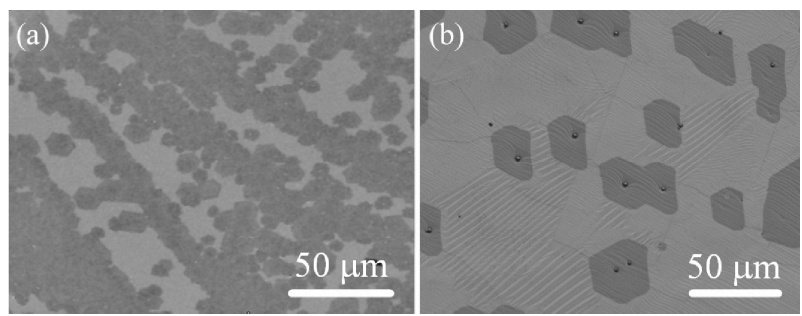

**Figure S2.** SEM images showing graphene nucleation on defect sites. (a) excessive nucleation along stripes due to cold rolling; (b) nucleation on abrasive particles embedded in the copper matrix due to improper polishing.

### 3) Enhancement of catalysis at higher temperatures

We examined the dependence of grain size on growth temperature (950 to 1070 °C) and the $H_2$ flow rate (0 to 10 sccm). Typical SEM images of graphene sample obtained at 950 °C, 1000 °C, 1050 °C, 1070 °C for 30 min are shown in Figure S3a, S3b, S3c and S3d, respectively. From the figures, higher temperature evidently favors the larger single crystal size, with the biggest grain size of about 200 μm obtained at 1050 °C and only 50 μm at 950 °C for 30 min. Meanwhile, the etching effect of $H_2$ manifests itself in the size control of graphene domain. Low $H_2$ flow rates yield dendrite-shaped domain due to high anisotropic growth, while high flow rate results in smaller domains (see Figure S3a and Figure S3b for comparison). It is worthy to mention that the attempt to make large graphene domain at 1070 °C failed (Figure S3d), probably because it is too close to the melting point of copper. The vibrant motion of copper atoms and dramatic evaporation of copper at this temperature may induce defects in graphene domains, which will serve as starting points for etching, resulting in many voids inside the single crystal domains.[4]

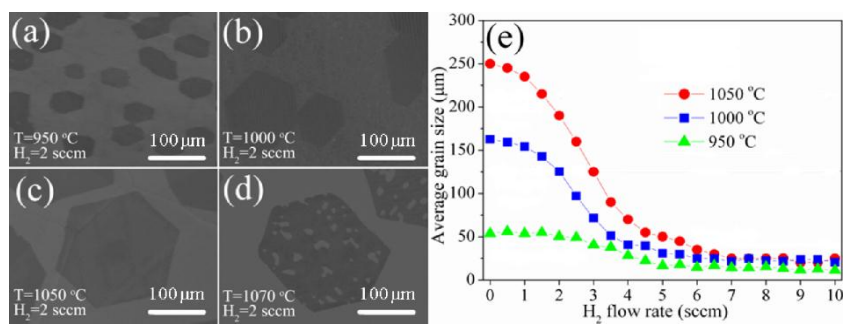

**Figure S3.** SEM images of graphene domains obtained at (a) 950 °C, (b) 1000 °C, (c) 1050 °C, (d) 1070 °C for 30 min and $T_p$=230 °C; (e) The size of graphene domains as a function of the $H_2$ flow rate at different growth temperature.

### 4) Stepwise growth of large graphene domains

Gradual increase of the $T_p$ and the total gas flow rate results in large graphene grains with smoothed edges (Figure 3d in the manuscript). Interesting to note is that if $T_p$ and flow rate were increased in a stepwise manner, stripe-like features will appear at the edge of the large graphene domain (Figure S4b).

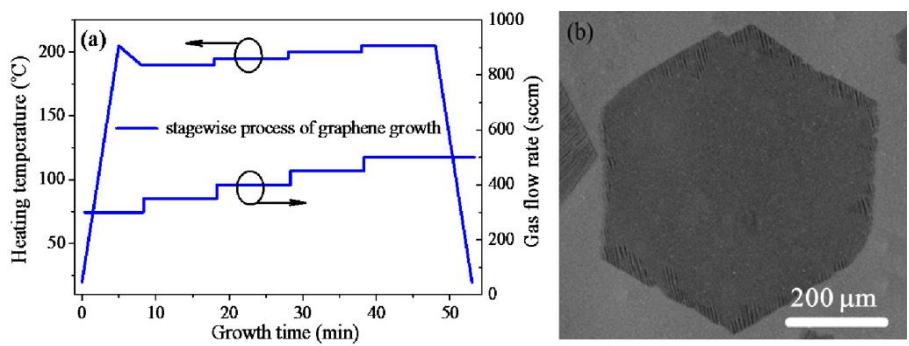

**Figure S4.** (a) APCVD growth of graphene at 1050 °C with stepwise increase of $T_p$ and total gas flow rate; (b) SEM image of hexagon-shaped graphene under stepwise growth process for 40 min.

### 5) Fabrication of the back gated field-effect transistors (FETs)

The graphene were transferred onto a 300 nm $SiO_2$/Si ( = 0.005 ·cm) substrate surface. Source and drain electrodes were prepared by standard photolithography using Cr/Au (5/50 nm) as contact metals. The highly doped silicon substrate was used as a back-gate electrode. The width of the source and the drain electrodes is about 5 m, and the distance between them is about 10 m. To obtain better contact between the graphene sheet and the Au/Cr electrodes, thermal annealing was performed in an $H_2$/Ar atmosphere at 300 °C for 20 min in a tube furnace. Optical image of the devices are shown in Figures S5.

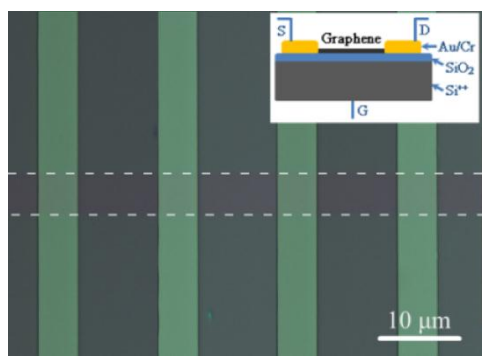

**Figure S5.** Optical image of fabricated graphene based field-effect transistor devices onto 300 nm $SiO_2$/Si substrates with patterned Au/Cr electrodes. The device is illustrated in the inset.